\documentclass[preprint,twocolumn,showpacs,prb,amsfonts,amsmath,amssymb,floatfix]{jpsj3}
\usepackage{hhline}
\usepackage{mathrsfs}
\usepackage{graphicx}
\usepackage{dcolumn}
\usepackage{bm}
\usepackage{color}
\usepackage{overcite}
\usepackage{cite}
\input{epsf}

\title{{\bfseries\itshape Ab initio} Derivation of Correlated Superatom Model for Potassium Loaded Zeolite A
}

\author{\name{Yoshiro \surname{Nohara}}\thanks{
Present address: Max-Planck-Institut f\"ur Festk\"orperforschung, Heisenbergstr. 1, D-70569 Stuttgart, Germany.}
, \name{Kazuma \surname{Nakamura}}$^1$, and \name{Ryotaro \surname{Arita}}$^{1,2}$}
\inst{\address{
Department of Physics, University of Tokyo, 7-3-1 Hongo Bunkyo-ku, Tokyo 113-0033, Japan \\
$^1$Department of Applied Physics, University of Tokyo, Tokyo 113-8656, Japan and\\
$^2$PRESTO, Japan Science and Technology Agency (JST), Kawaguchi, Saitama 332-0012, Japan}}

\abst{
We derive an effective low-energy Hamiltonian for potassium loaded zeolite A, a unique ferromagnet from non-magnetic elements.
We perform {\it ab initio} density functional calculations 
  and construct maximally localized Wannier functions for low-energy states made from potassium $s$ electrons.
The resulting Wannier orbitals, spreading widely in the alminosilicate cage, 
  are found to be the {\em superatomic} $s$ and $p$ orbitals in the confining potential formed by the host cage.
We then make a tight-binding model for these superatomic orbitals and introduce interaction parameters such as the Hubbard $U$.
After mean-field calculations for the effective model, 
  we find that {\it ab initio} spin density functional results are well reproduced by choosing appropriate sets of the interaction parameters.
The interaction parameters turn out to be as large as the band width, $\sim$ 0.5 eV, indicating the importance of electron correlation,
  and that the present system is an interesting analog of correlated multi-orbital transition metal oxides.
}
\kword{Zeolite, Magnetism, Superatom model, Density functional theory, Maximally localized Wannier functions}
\begin{document}
\maketitle 

\section{Introduction}
Electron correlation in materials, which often causes various non-trivial phenomena,
  has been one of the central issues in condensed matter physics.
While there are many kinds of fascinating strongly correlated systems
  such as organic molecular $p$-, transition metal $d$-, or heavy fermion $f$-electron systems,
  it is of great interest to consider the possibility of correlated $s$-electron systems.
Since valence $s$-electrons are usually itinerant, we expect that correlations in $s$-electron systems cannot be so significant.
However, recently, alkali-metal loaded zeolite is attracting broad interests as a unique exception;
  when clusters of alkali atoms are confined in the cage of alminosilicate,
  their kinetic energy is drastically suppressed and the system can exhibit a variety of correlation effects.

Although the unitcell of alkali-metal loaded zeolites is huge and complicated [typically it contains $O(100)$ atoms],
  the low-energy electronic structure of zeolites is surprisingly simple.
The host alminosilicate cage makes a gap of several eV, in which the guest alkali $s$-electrons form a few narrow bands around the Fermi level.
This situation is described by the so-called {\it superatom} picture,\cite{Nozue1,Nozue2,Arita}
  where each zeolite cage is regarded as a huge atom.
Here, the confining potential formed by the host framework and the guest alkali-metal $s$ electrons
  correspond to the atomic potential and valence electrons of the superatom, respectively.
Recent {\it ab initio} calculations have elucidated that the band dispersion around the Fermi level
  is well represented by a very simple tight-binding model,~\cite{Arita,SOD,SOD1,SOD2,SOD3,SOD4}
  suggesting that correlation effects in low-energy physics of zeolites 
  can be systematically described by the many-body superatom model.

As for {\it ab initio} derivation of many-body superatom models for zeolites,
  there have been several attempts for alkali-metal-loaded sodalite.~\cite{SOD,SOD1,SOD2,SOD3,SOD4}
Sodalite is one of the most prototypical zeolites
  made from the smallest building block of the alminosilicate cage (the so-called $\beta$-cage).
Each cage has only one valence $s$ electron, so that it can be viewed as a crystal of hydrogen-like superatoms.
In ref.~\citen{SOD}, a single-orbital extended Hubbard model for the superatomic $s$ electrons has been constructed,
  and the correlation strength was found to be much larger than the band width.

On the other hand, we also have many intriguing superatomic {\it multi-orbital} systems in the family of zeolites.
Potassium loaded zeolite A
  (K$_{16}$Si$_{12}$Al$_{12}$O$_{48}$, hereafter we call it K-LTA, where LTA is the code for the crystal structure of zeolite A)
  is a typical example, for which a spin-polarized ground state is realized.~\cite{Nozue1,Nozue2}
While the mechanism of spin-polarization has not been fully understood,
  there are several experimental\cite{Maniwa, Nakano-msr, SpinCant, Nakano-review}
  and theoretical\cite{Nohara,Arita} indications that orbital degrees of freedom of the superatomic $p$ orbital is important.
In this paper, we derive an effective multi-orbital Hamiltonian of K-LTA from first principles.
We estimate onsite potentials, transfer integrals, and interaction parameters in the Hamiltonian.
We show that the energy scale of the electron correlation, estimated as $\sim$ 0.5eV, is as large as the band width,
  so that the system is strongly correlated and can be regarded as a unique ``miniature'' of transition metal oxides.

The structure of the paper is as follows.
In \S\ 2, we first briefly describe the atomic configuration of K-LTA and give the level diagram of the low-energy states,
  which is useful to discuss the band structure of K-LTA in terms of the superatom picture.
In \S\ 3, we describe the detail of the method to evaluate the parameters in the effective Hamiltonian.
The basic idea is following: We first perform density functional calculation 
  and make maximally localized Wannier functions\cite{MaxLoc} for several bands around the Fermi level.
Using these Wannier functions, we then construct a tight-binding model and introduce interaction parameters such as the Hubbard $U$.
Choosing appropriate sets of the interaction parameters,
  we perform a mean-field calculation for the effective model and fit to the result of {\it ab initio} spin density functional calculation.
In \S\ 4, we present our results which indicate that the system is indeed classified as a strongly correlated system.
Section 5 is devoted to a summary and outlook.

\section{Level Diagram of Superatomic $p$-orbitals in Zeolite A}
\begin{figure}[htbp]
\vspace{0cm}
\begin{center}
\includegraphics[width=0.45\textwidth]{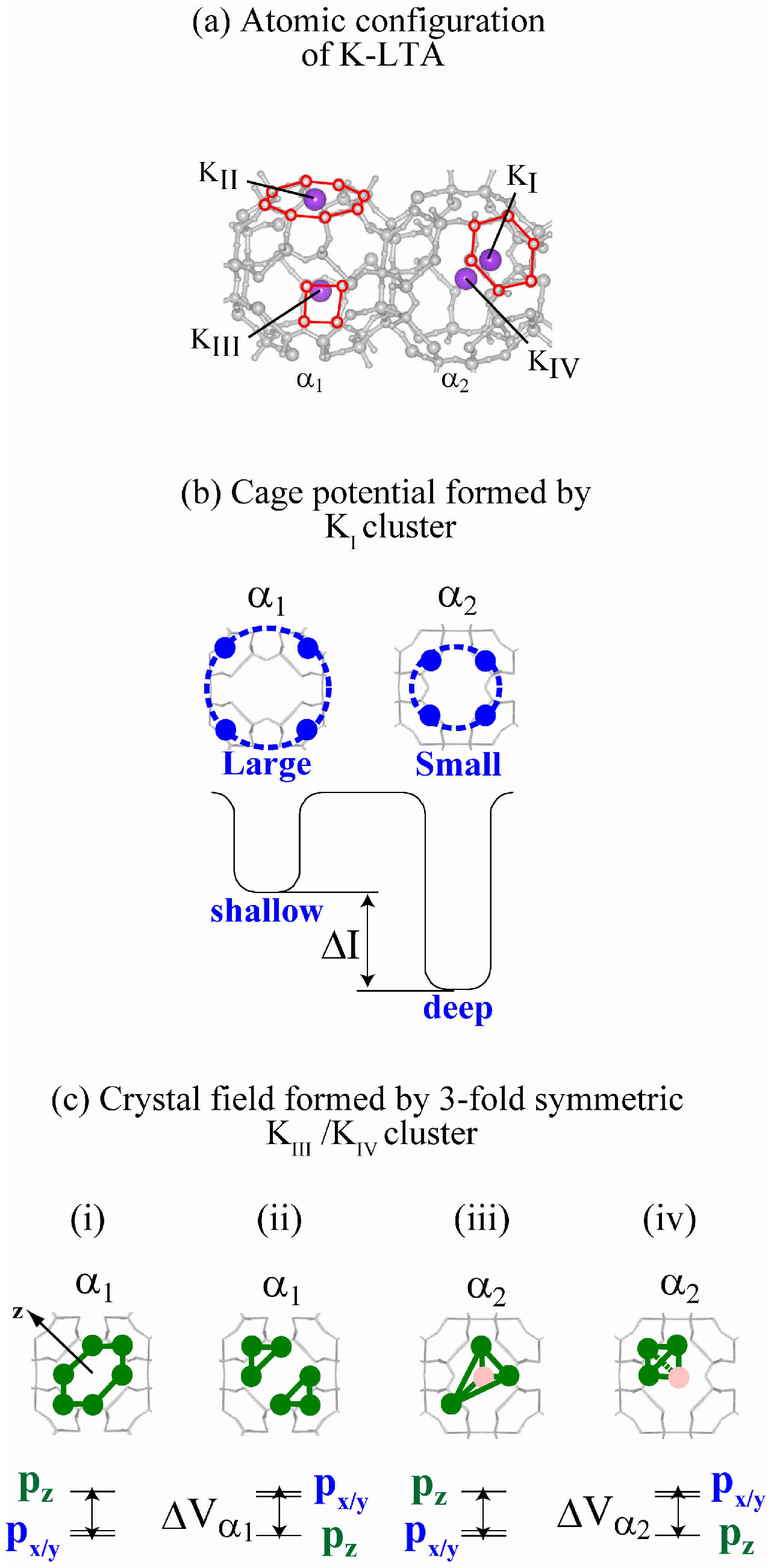}
\caption{
(Color online)
(a) Atomic configuration of K-LTA,
  (b) Difference in the cage-potentials ($\Delta I$) formed by K$_{{\rm I}}$ clusters [denoted by solid (blue) circles], and
  (c) Crystal-field splitting ($\Delta V_{\alpha_{i}}$) formed by clusters consisting of
  K$_{{\rm III}}$ and K$_{{\rm IV}}$ [denoted by solid dark gray (green) and light gray (pink) circles, respectively].
The $z$ axis is taken to be perpendicular to one of the six-membered rings in the $\alpha$ cage,
  and we assume that the system has three-fold symmetry along the $z$ axis.
Note that the sign and size of the crystal-field splitting depends on the atomic configuration of the cluster.
Four types of the K$_{{\rm III}}$ and K$_{{\rm IV}}$ configurations considered in the present study are depicted.
}
\label{fig_CPandCF3}
\end{center}
\end{figure}

Let us first look at the atomic configuration of K-LTA.
The system has two kinds of alminosilicate cage,
  the so-called $\alpha$ and $\beta$ cage (see Fig.~1 and related descriptions in ref.~\citen{Nohara}).
While the diameter of the $\beta$ cage is $\sim$ 7 \AA, the $\alpha$ cage is much larger ($\sim$ 11 \AA),
  so that the guest potassium cluster is accommodated only in the $\alpha$ cage.
In the unitcell, there are two inequivalent $\alpha$ and $\beta$ cages, which are denoted as $\alpha_1$, $\alpha_2$, $\beta_1$, and $\beta_2$.
Concerning the atomic positions of potassium atoms,
  neutron scattering measurements\cite{Ikeda} revealed
  that there are four kinds of sites: K$_{{\rm I}}$, K$_{{\rm II}}$, K$_{{\rm III}}$, and K$_{{\rm IV}}$ [Fig.~\ref{fig_CPandCF3}~(a)].
Here, the K$_{{\rm I}}$, K$_{{\rm II}}$, and K$_{{\rm III}}$ atoms
  sit around the center of the six-, eight-, and four-membered rings in the alminosilicate network, respectively,
  and the K$_{{\rm IV}}$ atom sits at the center of the $\alpha$ cage.
The sites of K$_{{\rm I}}$ and K$_{{\rm II}}$ are fully occupied, while the K$_{{\rm III}}$ and  K$_{{\rm IV}}$ sites are partially occupied.
Following the argument in ref.~\citen{Nohara}, we assume that the system has three-fold symmetry.

The low energy bands around the Fermi level are mainly made from the superatomic $p$ orbitals in the $\alpha$ cages.
When we discuss the superatomic $p$ levels, there are two key parameters.
One is the difference between the potentials of the two $\alpha$ cages, defined as
\begin{eqnarray}
\Delta I
\!=\!\min\{  \epsilon_{p_x \alpha_1},
           \!\epsilon_{p_y \alpha_1},
           \!\epsilon_{p_z \alpha_1}\!\}
\!-\!\min\{  \epsilon_{p_x \alpha_2}, 
           \!\epsilon_{p_y \alpha_2}, 
           \!\epsilon_{p_z \alpha_2}\!\}
\label{cage_pot} 
\end{eqnarray}
  with $\epsilon_{p_{\mu} \alpha_{i}}$ specifying the ionization potential of the superatomic $p_{\mu}$ orbital in the $\alpha_{i}$ cage.
Since the $\alpha_1$ cage accommodates more K$^{+}$ ions than the $\alpha_2$ cage,
  we naively expect that the potential of $\alpha_1$ is deeper than that of $\alpha_2$.
However, it should be noted here that the K$_{{\rm I}}$ atoms [denoted by solid (blue) circles in Fig.~\ref{fig_CPandCF3}~(b)]
  are distributed sparsely in the $\alpha_1$ cage than in the $\alpha_2$ cage,
  and the $\alpha_1$ cage does not have K$_{{\rm IV}}$ at the cage center [solid light gray (pink) circles in Fig.~\ref{fig_CPandCF3}~(c)].
These facts make, on the contrary to the above expectation,
  the $\alpha_1$-cage potential shallower than that of $\alpha_2$ [the schematic diagram in Fig.~\ref{fig_CPandCF3}~(b)].

The other important parameter is the crystal field splitting
  which is controlled by the atomic configuration of K$_{{\rm III}}$ [solid dark gray (green) circles in Fig.~\ref{fig_CPandCF3}~(c)].
For the K$_{{\rm III}}$ configuration spreading in the $xy$ plane [(i) and (iii) in the figure],
  the degenerated $p_x$ and $p_y$ levels become lower (lower schematic diagrams).
In contrast, for the configuration which extends in the $z$ direction [(ii) and (iv)], the crystal field lowers the $p_z$ level.
Note that the $\alpha_1$ cage contains 6 K$_{{\rm III}}$ atoms and the $\alpha_2$ cage does 3 K$_{{\rm III}}$ atoms.
Also note that the K$_{{\rm IV}}$ atom is accommodated only in the $\alpha_2$ cage.
The crystal-field splitting $\Delta V_{\alpha_{i}}$ in the $\alpha_{i}$ cage is defined as
\begin{eqnarray}
\Delta V_{\alpha_{i}}
 = \frac{\epsilon_{p_x \alpha_{i}} 
         +\epsilon_{p_y \alpha_{i}}}{2} 
 -       \epsilon_{p_z \alpha_{i}}.
\label{crystal_field} 
\end{eqnarray}

\begin{figure*}[htbp]
  \vspace{0cm}
\begin{center}
\includegraphics[width=0.80\textwidth]{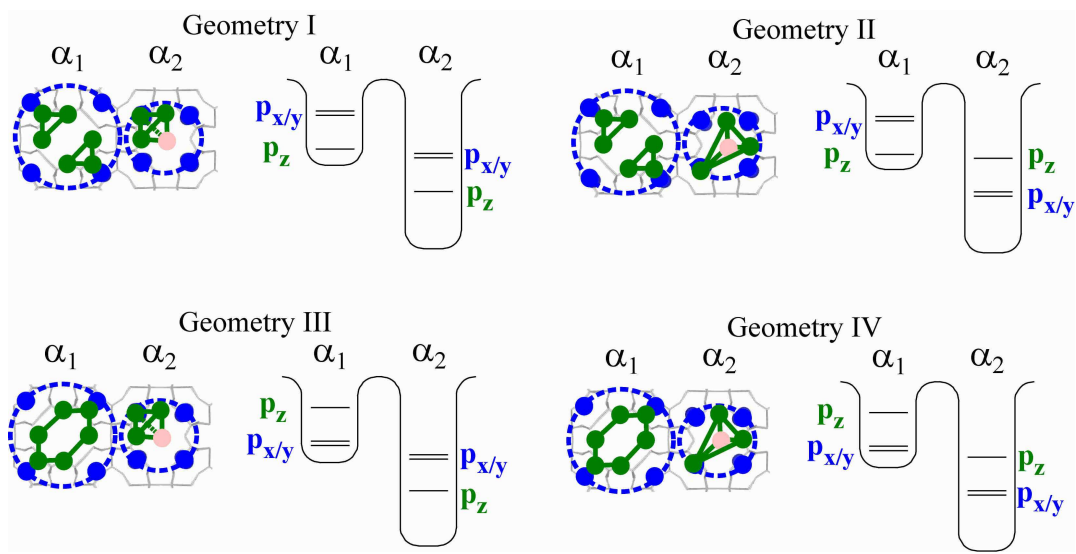}
\caption{
(Color online)
Geometries considered in the present study and their expected schematic level diagrams of the superatomic-$p$ orbitals.
}
\label{fig_CPandCF4}
\end{center}
\end{figure*}

In the previous study,\cite{Nohara}
  we performed spin density functional calculations for the four geometries in Fig.~\ref{fig_CPandCF4}, which we call geometries I-IV.
The figure also displays the schematic level diagrams with the superatomic picture.
It was found that the geometry I, whose total energy is the lowest among the four geometries,
  has the largest magnetic moment ($\sim 2 \mu_B$) in the $\alpha_2$ cage.
On the other hand, the magnetic moment is smallest for the geometry II.
This geometry dependence was simply discussed in terms of $\Delta I$ and \{$\Delta V_{\alpha_{i}}$\} in ref.~\citen{Nohara}.
While these are key parameters for the formation of local spin in the $\alpha$ cage,
  transfers between the neighboring $\alpha$ cages is also important to discuss the magnetism in the bulk.
We should also consider the degrees of freedom of the $\beta$ cage neglected in the previous analysis,
  mediating the hopping between the superatomic states in the $\alpha$ cage.
On top of that, careful analysis on the interplay between electron itinerancy and correlations are imperative.
Thus in the present study, we evaluate the interaction parameters such as Hubbard $U$ for the superatomic orbitals.

\section{Effective Model Construction}
Let us move on to the detail of the method to derive an effective model from first principles.
For evaluation of interaction parameters such as the Hubbard $U$, in the previous study for alkali-metal loaded sodalite,\cite{SOD}
  we used the constrained random phase approximation.~\cite{cRPA1,cRPA2}
However, since the system size of K-LTA is much larger than that of sodalite and the numerical cost is so expensive,
  here we take a different approach.

First, we introduce the following model Hamiltonian, 
\begin{eqnarray}
&\cal{H}& \equiv {\cal{H}}_0 + {\cal{H}_{{\rm int}}}, \label{ham_tot} \\
&{\cal{H}}_0&\!\equiv\!\sum_{i,m,\sigma}\!\epsilon_{im} n_{im\sigma}\!+\!\sum_{\sigma,(im)\ne(jm')}\!t_{im jm'}  c^{\dagger}_{im\sigma} c_{jm'\sigma}, \label{ham_0} \\
&{\cal{H}}_{{\rm int}}& 
\equiv \sum_{i,m}U_{im} n_{im\uparrow} n_{im\downarrow} 
+      \sum_{i,m<m',\sigma} U'_{imm'} n_{im\sigma}n_{im'\bar{\sigma}} \nonumber \\
& & +  \sum_{i,m<m',\sigma}(U'_{imm'}-J_{imm'}) n_{im\sigma} n_{im'\sigma}, 
\label{ham_u} 
\end{eqnarray}
  where $c_{im\sigma}^\dagger (c_{im\sigma})$ creates (annihilates) an electron with spin $\sigma$ in orbital $m$ at site $i$,
  and $n_{im\sigma}\equiv c_{im\sigma}^\dagger c_{im\sigma}$.
The term ${\cal{H}}_0$ describes the kinetic-energy part of the effective model,
  where $\epsilon_{im}$ and $t_{im jm'}$ are the ionization potential and transfer, respectively.
In this model, we take into account of not only the superatom $p$ states in the $\alpha$ cage
  but also the superatom $s$ states in the $\beta$ cage, which lie just above the Fermi level as explained in the next section.
The term $\cal{H}_{{\rm int}}$ is the interaction part, i.e.,
  the intraorbital ($U_{im}$) and interorbital ($U'_{imm'}$) Coulomb interactions and the Ising component of Hund's coupling ($J_{imm'}$).
Since the standard spin density functional calculation is not SU(2) symmetric,
  we ignore the spin-flip component of the Hund's coupling and the pair-hopping interaction.
For simplicity, we also ignore orbital dependence of interaction parameters in the $\alpha$ cage;
  common $U_{\alpha}$ is used for any $m$ and common $U'_{\alpha}$ and $J_{\alpha}$ are used for any combination of $m$ and $m'$.
For the $\beta$ cage, there is no orbital degree of freedom.

Now, let us consider to derive the interaction parameters \{$U_{\alpha}$, $U_{\beta}$, $U'_{\alpha}$, and $J_{\alpha}$\}.
These parameters are determined by the fitting procedure in such that
  the Hartree-like mean field solution of the model reproduces
  both of {\em ab initio} non-magnetic and spin-polarized density functional results based on the generalized gradient approximation (GGA).

In order to relate the mean field in the model calculation and the density functional calculation, we introduce the following two terms.
One is a term to scale the self interaction (SI) and the other is a term which guarantees that
  the non-magnetic model calculation gives the same result as the non-magnetic density functional calculation.
For the first issue, as is well known, GGA includes the SI effect.~\cite{LDAreview}
On the other hand, the Hubbard model defined in eq.~(\ref{ham_tot}) does not have SI;
  the intra-orbital interaction between electrons with the same spin are excluded in the model Hamiltonian.
Therefore, to mimic the GGA situation in the model calculation, we need to introduce the SI term as
\begin{eqnarray}
{\cal{H}}_{\rm SI}[\rho] 
\equiv 
\sum_{i,m,\sigma} S_i U_{i} \rho_{im\sigma m\sigma} c^{\dagger}_{im\sigma} c_{im\sigma}, 
\end{eqnarray}
  where $\rho_{im\sigma n\sigma'}=\langle c^{\dagger}_{in\sigma'} c_{im\sigma} \rangle$ is the density matrix
  and the parameter $S_{i}$ scales the size of SI.
It should be noted that ${\cal{H}}_{\rm SI}[\rho]$ is a functional of the diagonal part of the density matrix.
With this treatment, we can estimate the size of SI in GGA by ${\cal H}_{{\rm SI}}[\rho_{{\rm GGA}}]$,
  where $\rho_{\rm GGA}$ is the density matrix obtained from the {\em ab initio} non-magnetic GGA calculations.
For the second issue, we introduce $\bar{{\cal H}}[\rho_{\rm GGA}]$,
  where $\bar{{\cal H}}[\rho]$ denotes the Hartree approximation for ${\cal H}_{{\rm int}}$, being a functional of the density matrix $\rho$.
The $\bar{{\cal H}}[\rho_{\rm GGA}]$ term is evaluated with the density obtained by non-magnetic GGA.

Thus the one-body part ${\cal H}_0$ in eq.~(\ref{ham_0}) is defined as 
\begin{eqnarray*}
{\cal H}_0 
\equiv \cal{H}_{{\rm GGA}} 
- \cal{H}_{\rm SI}[\rho_{{\rm GGA}}] 
- {\bar{\cal{H}}[\rho_{\rm GGA}]},  
\end{eqnarray*}
  where $\cal{H}_{{\rm GGA}}$ is the tight-binding representation of the non-magnetic GGA Kohn-Sham Hamiltonian
  with $\epsilon_{im}^{{\rm GGA}}$ and $t_{im jm'}^{{\rm GGA}}$ in diagonal and off-diagonal parts, respectively.
We use the maximally localized Wannier functions for the basis functions for those matrix elements.
More precisely, the matrix elements of ${\cal H}_0$ are calculated as
\begin{eqnarray}
\epsilon_{im}
&=& \epsilon_{im}^{\rm GGA} 
 -  \Biggl[ \frac{U_i (S_i+1)}{2} \Biggr] \rho^{{\rm GGA}}_{imm} \nonumber \\
&-& \Biggl( U'_i - \frac{J_i}{2} \Biggr) 
    ( N^{{\rm GGA}}_i - \rho^{{\rm GGA}}_{im m} ) \label{IP}
\end{eqnarray}
  and  
\begin{eqnarray}
t_{im jm'} = t_{im jm'}^{\rm GGA}. \label{transfer}
\end{eqnarray}
Here, $\rho_{imm}^{{\rm GGA}}$ is the matrix element of the non-magnetic GGA density matrix
  and $N^{{\rm GGA}}_{i}\!\equiv\!\sum_{m} \rho^{{\rm GGA}}_{imm}$.
Note that transfers in ${\cal H}_0$ is the same as those in ${\cal H}_{\rm GGA}$,
  since ${\cal H}_{\rm SI}$ and $\bar{\cal H}$ are local in real space.

Now, the mean-field Hamiltonian ${\cal H}_{{\rm MF}} [\rho]$ for ${\cal H}$ reads as 
\begin{eqnarray}
  {\cal H}_{\rm MF} [\rho]
= {\cal H}_0 
+  \bar{{\cal H}} [\rho]
+ {\cal H}_{\rm SI} [\rho]. 
\label{ham_mf}
\end{eqnarray}
Note that the result of non-magnetic GGA is exactly reproduced as $\cal{H}_{\rm MF} [\rho_{\rm GGA}]$ in the present framework. We then consider to solve $\cal{H}_{\rm MF}$ by spin-polarized self-consistent Hartree scheme. We can estimate the interaction and self-interaction parameters, $U_{\alpha}$, $U_{\beta}$, $U'_{\alpha}$, $J_{\alpha}$, $S_{\alpha}$, and $S_{\beta}$, by fitting the model band dispersion to the {\em ab initio} spin density functional results. 

\vspace{1cm} 
\section{Results}
Using the {\it Tokyo Ab initio Program Package},\cite{TAPP} we first performed density functional calculations with the GGA exchange-correlation functional.\cite{PBE96} We used a plane-wave basis set with the ultrasoft pseudopotentials.\cite{USPP} The energy cutoff was set to 36 Ry for wavefunction and 144 Ry for spin and charge densities and 4$\times$4$\times$4 ${\bf k}$-mesh was adopted. Maximally localized Wannier function with the ultrasoft pseudopotential was constructed following the recipe of ref.~\citen{Ferretti}. 

\subsection{Maximally localized Wannier functions}
\begin{figure*}[htbp]
  \vspace{0cm}
  \begin{center}
  \includegraphics[width=0.80\textwidth]{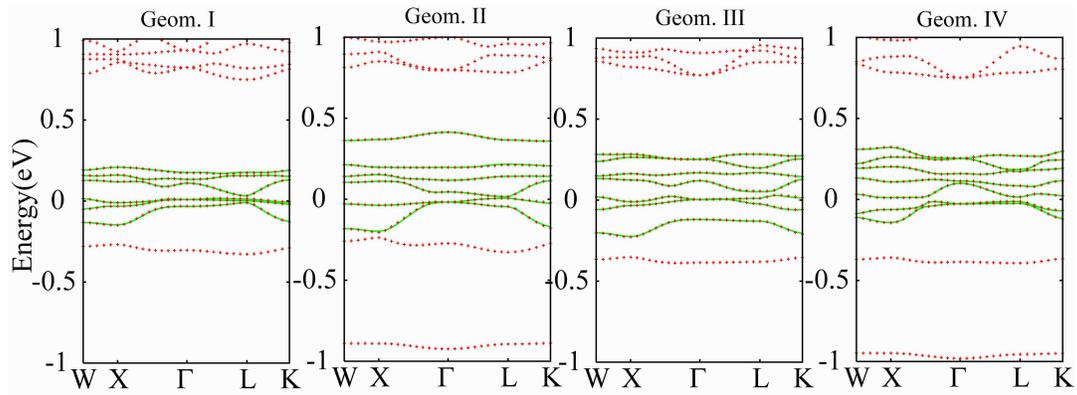}
  \caption{
(Color online)
GGA band dispersions [(red) dotted lines]
  and interpolated band dispersions for the maximally localized Wannier functions [(green) solid lines] for geometry I-IV.
Energy is referenced to the Fermi level.
}
  \label{fig_wannier_fitting}
  \end{center}
\end{figure*}

The GGA band structures for the geometries I-IV are shown in Fig. \ref{fig_wannier_fitting} as (red) dotted lines.
We see that the low-energy band structures depend sensitively on the geometry.
Especially, the bandwidth is rather different from each other.
From the six (seven) bands around the Fermi level,
  we constructed the maximally localized Wannier functions for the geometry I and II (III and IV),
  which are employed as bases of the effective Hamiltonian in eq.~(\ref{ham_tot}).
We estimated tight-binding parameters for the geometries I-IV by calculating matrix elements of ${\cal H}_{\rm GGA}$ in the Wannier basis.
In Fig. \ref{fig_wannier_fitting},
  we plot interpolated band dispersions [(green) solid lines],
  from which we see that original {\em ab initio} bands are well reproduced by the Wannier interpolation.
The band around $-$0.4$\sim$$-$0.3 eV is formed by the superatomic $s$ orbital in the $\alpha$ cage,
  which is not included in the effective model.

As shown in detail later (Fig. \ref{fig_fitting}), the $s$ orbital in the $\alpha$ cage is fully occupied
  and its exchange splitting is negligibly small. Thus, for simplicity, we do not consider explicitly
  this degree of freedom in the low-energy model.
It should be noted that the $s$ state in the $\alpha$ cage has small transfer integrals between the surrounding sites 
  and tiny onsite hybridizations
  with the $p$ states, because it is well localized in the cage and the confining potential is almost spherical.
Moreover, the $s$ state should have larger $U$ and $S$ than the $p$ states, so that there should be a significant difference
  in the ionization potential, $\epsilon_{im}$ [eq.~(\ref{IP})], for the $s$ and $p$ states.
Therefore, the $s$ state is expected to be irrelevant for the magnetic properties.

\begin{figure}[htbp]
  \vspace{0cm}
  \begin{center}
  \includegraphics[width=0.45\textwidth]{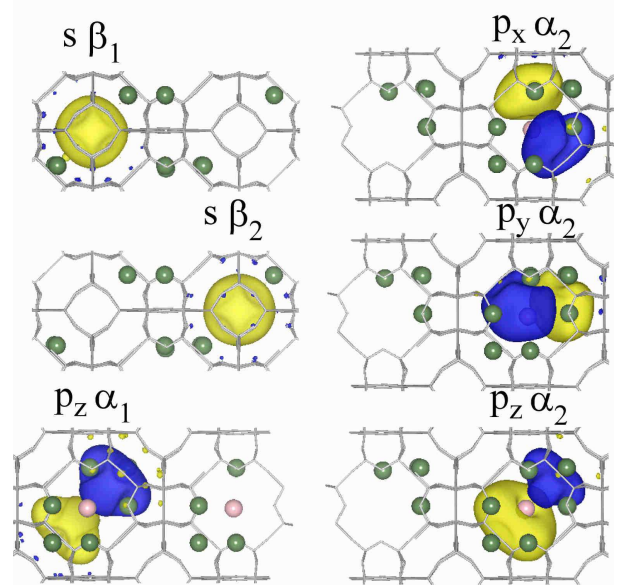}
  \caption{
(Color online)
Isosurface contours of maximally localized Wannier functions for superatomic-$p$ (-$s$) orbitals
  in the $\alpha$ ($\beta$) cage of the geometry I.
The amplitudes of the contour surface are 0.015 [light gray (yellow)] and $-$0.015 [dark gray (blue)] in the atomic unit.
Dark gray (green) and light gray (pink) spheres represent potassium K$_{\rm III}$ and K$_{\rm IV}$, respectively.
}
  \label{fig_wannier}
  \end{center}
\end{figure}

In Fig. \ref{fig_wannier}, we display the isosurface contours of the plane-wave part of the Wannier functions for the geometry I.
In this geometry,
  there are six bases ($s\beta_1$, $s\beta_2$, $p_z\alpha_1$, $p_z\alpha_2$, $p_y\alpha_2$, and $p_x\alpha_2$),
  and the $p_x\alpha_1$ and $p_y\alpha_1$ states are not included in the effective model,
  because the latter two exist far above the Fermi level by, at least, $\sim$0.5 eV.
Similarly, for the geometries II-IV, the high-energy $p_{\mu}\alpha_1$ states were not included in the effective models.
Table \ref{tab_wannier} lists our calculated spatial spreads of the Wannier functions, $\Omega_{mi}$, for all the geometries.
As is indicated in the table,
  the size of the Wannier functions is as large as the diameter of the host cage
  ($\sim$11 \AA\ for the $\alpha$ cage and $\sim$7 \AA\ for the $\beta$ cage)
  and $\Omega_{\beta}$ $\le$ $\Omega_{m\alpha}$, 
  so that the onsite interactions in the $\beta$ cages should be larger than those in the $\alpha$ cages.

\begin{table}[htbp] 
\caption{
Spatial spread of maximally localized Wannier functions for superatomic states, $\Omega$,
  and self-consistently determined ionization potentials, $\epsilon$ defined by eq.~(\ref{IP}).
The difference in cage potentials $\Delta I$ [eq.~(\ref{cage_pot})]
  and the crystal-field splitting in the $\alpha_2$ cage $\Delta V_{\alpha_2}$ [eq.~(\ref{crystal_field})] are also listed.
The unit is given as \AA\ for $\Omega$ and eV for $\epsilon$, $\Delta I$, and $\Delta V_{\alpha_2}$.}

\centering 
{\scriptsize 
\begin{tabular}{c@{\ }r@{\ }r@{\ }r@{\ }r@{\ }r@{\ }r@{\ }r@{\ }r@{\ }r@{\ }r@{\ }r} 
 \hline \hline \\ [-5pt]  
       & \multicolumn{2}{c}{Geom I}   &      
       & \multicolumn{2}{c}{Geom II}  &
       & \multicolumn{2}{c}{Geom III} &
       & \multicolumn{2}{c}{Geom IV}  
\\ [2pt] 
\hline \\ [-5pt] 
                 & $\Omega$\ \ \ & $\epsilon$\ \ \ &    
                 & $\Omega$\ \ \ & $\epsilon$\ \ \ & 
                 & $\Omega$\ \ \ & $\epsilon$\ \ \ & 
                 & $\Omega$\ \ \ & $\epsilon$\ \ \ 
\\ [2pt] \hline \\ [-5pt]
$s\beta_1$       & 4.42 &$-$0.05&    
                 & 4.65 & 0.04& 
                 & 4.34 & 0.04& 
                 & 4.44 & 0.02\\          
$s\beta_2$       & 4.36 & 0.06&   
                 & 4.45 & 0.14& 
                 & 4.36 & 0.11& 
                 & 4.44 & 0.09\\      
$p_z\alpha_1$    & 5.22 &$-$0.20&  
                 & 5.54 &$-$0.07& 
                 & -    & -   &
                 & -    & -   \\          
$p_y\alpha_1$    & -    & -   &  
                 & -    & -   & 
                 & 5.31 & 0.08& 
                 & 5.33 & 0.11\\  
$p_x\alpha_1$    & -    & -   &  
                 & -    & -   & 
                 & 5.31 & 0.08& 
                 & 5.33 & 0.11\\ 
$p_z\alpha_2$    & 5.15 &$-$0.78&  
                 & 5.11 &$-$0.03& 
                 & 5.39 &$-$0.48& 
                 & 5.43 &$-$0.25\\ 
$p_y\alpha_2$    & 5.73 &$-$0.68&  
                 & 5.95 &$-$0.39& 
                 & 5.37 &$-$0.34& 
                 & 5.26 &$-$0.36\\ 
$p_x\alpha_2$    & 5.74 &$-$0.68&  
                 & 5.94 &$-$0.39& 
                 & 5.37 &$-$0.34& 
                 & 5.26 &$-$0.36\\ 
\hline \\ [-4pt] 
\multicolumn{1}{l}{$\Delta I$} 
       & \multicolumn{2}{c}{\ \ 0.58} &      
       & \multicolumn{2}{c}{\ \ 0.32} &
       & \multicolumn{2}{c}{\ \ 0.56} &
       & \multicolumn{2}{c}{\ \ 0.47} \\ 
\multicolumn{1}{l}{$\Delta V_{\alpha_2}$} 
       & \multicolumn{2}{c}{\ \ 0.10} &      
       & \multicolumn{2}{c}{$-$0.36}  &
       & \multicolumn{2}{c}{\ \ 0.15} &
       & \multicolumn{2}{c}{$-$0.12}  \\ [2pt]
\hline \hline \\ 
\end{tabular}
}  
\label{tab_wannier} 
\end{table}

\subsection{Effective low-energy Hamiltonian}
Let us move on to the details of the effective model derived with the present scheme.
First, the transfer integrals can be read from the table of the matrix elements of ${\cal H}_{\mathrm{GGA}}$ in the Wannier basis [eq.~(\ref{transfer})].
The nearest-neighbor hoppings between the $\alpha$ and $\beta$ cages
  are $|t_{p_{x/y}\alpha s\beta}|=0.01\sim 0.03$eV and $|t_{p_{ z }\alpha s\beta}|=0.02\sim 0.04$eV.
On the other hand, the transfers between neighboring $\alpha$ cages are $|t_{p_{ z }\alpha p_{ z }\alpha}|=0.01\sim 0.02$eV,
  $|t_{p_{ z }\alpha p_{x/y}\alpha}|=0.02\sim 0.03$eV, and $|t_{p_{x/y}\alpha p_{x/y}\alpha}|=0.03\sim 0.04$eV.
The distant transfers are negligibly small.
We remark that, among the transfers between $\alpha$ cages, the $p_z$-$p_z$ hopping is smaller than those of other $p$ states.
This is because the $p_z$ orbital is not pointing at the neighboring $\alpha$ cages.
Thus, the $p_z$ states tend to be localized and to have large magnetic moments.

In order to estimate \{$U_{\alpha}$, $U_{\beta}$, $U'_{\alpha}$, $J_{\alpha}$, $S_{\alpha}$, $S_{\beta}$\},
  we performed spin-polarized self-consistent calculations for ${\cal H}_{\rm MF}$ in eq.~(\ref{ham_mf}),
  and looked for the parameter set which reproduces the band dispersion of {\it ab initio} spin density functional calculations.
First, we searched for the parameter set
  for which the difference between the GGA band dispersion of the geometry I
  and that of the model calculation
  along the symmetry line W$\rightarrow$X$\rightarrow \Gamma \rightarrow $L$ \rightarrow$ K in the Brillouin zone is minimized.
We assumed that $U_\alpha < U_\beta <0.6$eV, $S_\alpha < S_\beta < 1$, $U'_\alpha/U_\alpha<1$,
  and $J_\alpha/U_\alpha<0.2$ with the realistic parameter range.~\cite{SOD}
To understand and explain this system by a minimum set of parameters,
  we then introduced a scaling factor $x$ for \{$U_\alpha, U_\beta, U'_\alpha, J_\alpha$\}
  and minimize the band-dispersion difference in GGA and model for the geometry II, III, and IV.
We found that the GGA band structures for these three geometries are successfully reproduced for $x=0.6$, $S_\beta=0.7$, and $S_\alpha=0.1$.
These parameters show that Geom. II-IV have smaller interaction parameters than Geom. I.
This is consistent with the fact
  that the band widths in Geom. II-IV are broader than that in Geom. I as shown in Fig. \ref{fig_wannier_fitting}.

\begin{figure*}[htbp]
  \vspace{0cm}
  \begin{center}
  \includegraphics[width=0.80\textwidth]{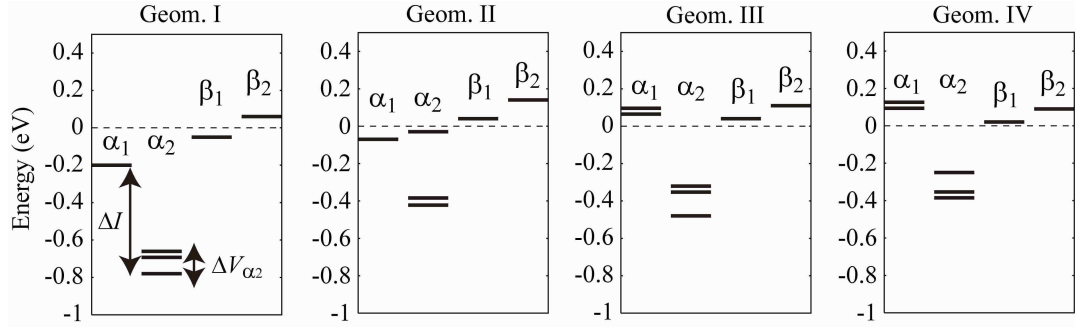}
  \caption{
Level diagrams for the superatomic $p$ levels in the $\alpha$ cages and the $s$ levels in the $\beta$ cages for the geometries I-IV.
Energy zero is the Fermi level (dotted line).
The difference between the cage potentials of $\alpha_1$ and $\alpha_2$,
  $\Delta I$ in eq.~(\ref{cage_pot}), and the crystal-field splitting in $\alpha_2$ cage, $\Delta V_{\alpha_2}$ in eq.~(\ref{crystal_field}), 
  are also depicted in the geometry I.
}
  \label{level}
  \end{center}
\end{figure*}

Before presenting the resulting interaction parameters,
  we summarize in Table~\ref{tab_wannier} our calculated self-consistent non-magnetic ionization potentials, $\epsilon$ in eq.~(\ref{IP}).
Using the $\epsilon$ data,
  we estimated the effective cage-potential differences $\Delta I$ in eq.~(\ref{cage_pot})
  and the crystal field splitting in the $\alpha_2$ cage $\Delta V_{\alpha_{2}}$ in eq.~(\ref{crystal_field}),
  which are given in the bottom two rows in Table~\ref{tab_wannier}.~\cite{Va1}
It should be noted here that these one-body quantities have appreciable geometrical dependence,
  suggesting that slight changes in the potassium configurations can cause substantial changes in the low-energy electronic structure of K-LTA.
According to Table~\ref{tab_wannier}, we draw in Fig.~\ref{level} the level diagram of the superatomic states in the $\alpha$ and $\beta$ cages.
We can see that the $\alpha_2$-cage potential is significantly deep compared to the other cage potentials 
  and thus will form the trapping potentials for the low-energy electrons.

\begin{figure*}[htbp]
  \vspace{0cm}
  \begin{center}
  \includegraphics[width=0.80\textwidth]{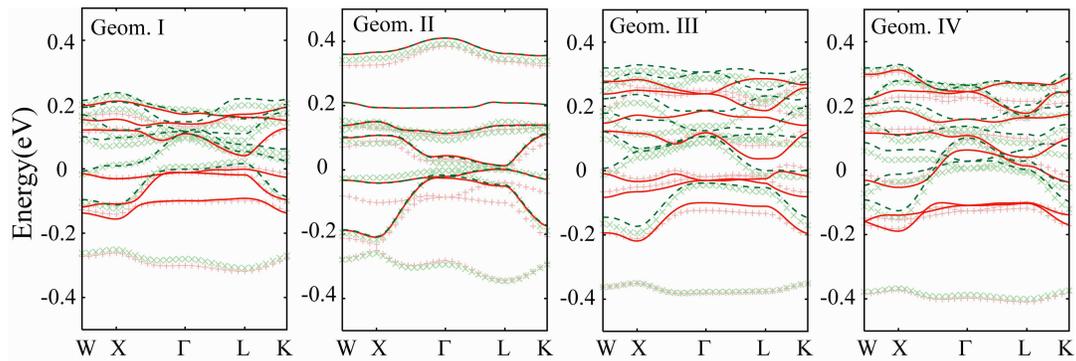}
  \caption{
(Color online)
Band dispersions of majority [(red) +] and minority [(green) $\times$] spins obtained by the spin-polarized GGA,
  and those of majority [(red) solid line] and minority spins [(green) dashed line] calculated by the model calculation for the geometries I-IV.
Energy is reference to the Fermi level.
}
  \label{fig_fitting}
  \end{center}
\end{figure*}

We next show in Fig.~\ref{fig_fitting}
  our calculated spin-polarized model band dispersion of the majority spin [(red) solid lines] and minority spin [(green) dashed lines],
  comparing with those of the {\it ab initio} results [majority (+) and minority ($\times$)].
Table \ref{tab_etotlmom} compares the resulting local magnetic moments in the model and {\it ab initio} calculations.
In spite of the limited number of the free parameters, the {\em ab initio} band dispersion and moment size are well reproduced.~\cite{Nohara}
The magnetic moments are basically formed in the $\alpha_2$ cage, being consistent with the level diagram in Fig.~\ref{level}.
The transfers cause the antiferromagnetic correlation between the $\alpha_1$ and $\alpha_2$ cages.

The interaction parameters estimated by the present fitting are summarized in Table \ref{tab_interaction}.
We see that the resulting $U$ is 0.24-0.5 eV and is always about ten times larger than the transfer integrals as 0.01-0.04 eV,
  thus indicating that the system resides in moderately/strongly correlation regime.
We note that interaction parameters depend sensitively on the atomic configuration;
  they tend to be large (small) for the geometry I (II-IV).~\cite{note_int}
It is also interesting to note that the self-interaction parameter $S_i$ tends to be larger when the interaction parameters are large.
This is a reasonable trend,~\cite{LDAreview} because self-interaction is generally large for localized states,
  and localized states usually have strong onsite electron correlation.
In the present study, the $S_i$ value is estimated by parameter fitting to the spin polarized GGA calculations.
Therefore, the present $S_i$ parameter qualitatively measures the localization of the spin-polarized wavefunctions,
  but not that of the paramagnetic wavefunctions.
\begin{table}[htbp] 
\caption{Local magnetic moment ($M$, in the unit of $\mu_B$) in the {\em ab initio} and the model calculation for K-LTA.}
 
\centering 
{\scriptsize 
\begin{tabular}{c@{}r@{}r@{}r@{}r@{}r@{}r@{}r@{}r@{}r@{}r@{}r} 
 \hline \hline \\ [-5pt]  
       & \multicolumn{2}{c}{Geom I}   &      
       & \multicolumn{2}{c}{Geom II}  &
       & \multicolumn{2}{c}{Geom III} &
       & \multicolumn{2}{c}{Geom IV}  
\\ [2pt] 
\hline \\ [-5pt] 
                 & {\em ab initio\ } & {Model\ } &    
                 & {\em ab initio\ } & {Model\ } & 
                 & {\em ab initio\ } & {Model\ } & 
                 & {\em ab initio\ } & {Model\ }  
\\ [2pt] \hline \\ [-5pt]
$M_{\beta_1}$    &  -  & 0.00 &    
                 &  -  & 0.00 & 
                 &  -  & 0.08 & 
                 &  -  & $-$0.09 \\          
$M_{\beta_2}$    &  -  & 0.02 &   
                 &  -  & 0.00 & 
                 &  -  & 0.01 & 
                 &  -  & $-$0.03 \\      
$M_{\alpha_1}$   &$-$0.02& $-$0.20 &  
                 &$-$0.20& $-$0.02 & 
                 &   0.06& 0.20 &
                 &$-$0.16& $-$0.07 \\          
$M_{\alpha_2}$   & 1.92  & 2.19 &  
                 & 0.33  & 0.04 & 
                 & 1.36  & 1.69 & 
                 & 1.00  & 1.61 \\   
[2pt]   
\hline \hline \\ 
\end{tabular}
}  
\label{tab_etotlmom}
\end{table}
\begin{table}[htbp]
\caption{Estimated interaction parameters ($U$, $U'$, $J$) and self-interaction parameter ($S$) in K-LTA}.
\label{tab_interaction}
\begin{center}
\begin{tabular}{c@{\ \ \ \ }r@{\ \ \ \ }r@{\ \ \ \ }r@{\ \ \ \ }r}
\hline \hline
                  & Geom I  & Geom II  & Geom III & Geom IV \\ \hline
$U_{\beta}$(eV)   & 0.5     & 0.3      & 0.3      & 0.3     \\
$U_{\alpha}$(eV)  & 0.4     & 0.24     & 0.24     & 0.24    \\
$J_{\alpha}$(eV)  & 0.03    & 0.018    & 0.018    & 0.018   \\
$U'_{\alpha}$(eV) & 0.24    & 0.144    & 0.144    & 0.144   \\
$S_{\beta}$       & 0.9     & 0.7      & 0.7      & 0.7     \\
$S_{\alpha}$      & 0.7     & 0.1      & 0.1      & 0.1     \\
\hline \hline
\end{tabular}
\end{center}
\end{table}

Finally, let us comment on the controversial situation in the experiment.
Experimentally, two kinds of possible magnetic structures have been proposed for ferromagnetism of K-LTA.
One is the ferrimagnetic model,\cite{Maniwa} and the other is the spin-canted antiferromagnetic model.\cite{SpinCant}
In the former, the system consists of two sublattices
  with large ($\sim$ 2.8 $\mu_{{\rm B}}$) and small (almost 0.0 $\mu_{{\rm B}}$) magnetic moment,
  while, in the latter, every $\alpha$ cage has a moment of 1 $\mu_B$ and the spins are canted
  in the low temperature by the spin-orbit interaction.
These two magnetic structures seem to agree with the results for the geometries I and IV, respectively,~\cite{Nohara}
  although the size of the moments is relatively smaller than the experimental values.
It is interesting to note that the geometries I and IV are energetically stable
  (slightly lower in the geometry I by $\sim$0.1 eV).~\cite{Nohara}
The present results seem to support the former proposal,
  but we should emphasize that K-LTA resides in moderately or strongly correlation regime
  so that careful analyses with proper consideration of the correlation effect are required
  to clarify the mechanism of the spin-polarization in real K-LTA.
Indeed, strongly correlated multi-orbital systems generally have rich phase diagrams.
For example, a subtle competition between the interaction parameters and the tight-binding parameters
  can drastically change the magnetic properties of the system.
Elaborated many-body analyses for K-LTA are indeed a fascinating future problem.


\section{Summary and Outlook}

We have derived an effective low-energy Hamiltonian for K-LTA.
First, we constructed a tight-binding model from GGA using the maximally localized Wannier functions.
We then introduced the interaction and self-interaction parameters, and solved the model self-consistently.
With appropriate sets of these parameters, we have succeeded in reproducing the result of spin density functional calculation.
We have found that the system reside in the strong coupling regime in that the bandwidth and electron correlation have the same energy scale.
We have also shown that the details of the potassium-cluster configurations
  crucially affects both the one-body and two-body parts of the effective Hamiltonian of K-LTA.

While the present method of evaluating the interaction parameters
  has a great advantage in that it can be applied to huge systems such as K-LTA,
  it does not work for systems for which spin density functional calculation gives a non-magnetic ground state.
Since the size of magnetic moment tends to be underestimated in GGA,
  the present method is expected to give the lower bound of the interaction parameters.
It is indeed an interesting future challenge
  to apply other {\it ab initio} methods to evaluate interaction parameters
  such as constrained density functional theory (ref.~\citen{cLDA})
  or constrained random-phase approximation (refs.~\citen{cRPA1} and \citen{cRPA2}) to K-LTA.
While formidable numerical cost will be required,
  comprehensive model construction by different methods is important to establish the low-energy correlation physics in the zeolite system.

\begin{acknowledgments}
We thank Professor Yasuo Nozue and Takehito Nakano for fruitful discussions.
This work was supported by Grants-in-Aid for Scientific Research (No.~19051016, 22740215, 22104010, and 23110708)
  and the Next Generation Super Computing Project from MEXT, Japan,
  and ``Funding Program for World-Leading Innovative R\&D on Science and Technology (FIRST program)'' JSPS,
  Japan and the JST PRESTO program, Japan.
Numerical calculations were done at the ISSP supercomputer center,
  and the Supercomputing Division, Information Technology Center, University of Tokyo.
\end{acknowledgments}


\begin{thebibliography}{99}
\bibitem{Va1}
The crystal field splitting in the $\alpha_1$ cage $\Delta V_{\alpha_1}$ cannot be defined,
  because the highest $p$ level is not included in the effective Hamiltonian.
In fact, $|\Delta V_{\alpha_1}|$ is expected to be much larger than $|\Delta V_{\alpha_2}|$,
  since the number of K$_{\rm III}$ cations in the $\alpha_1$ cage is two times larger than the $\alpha_2$ cage
  (6 for $\alpha_1$ and 3 for $\alpha_2$).
\bibitem{note_int}
In the present scheme, interaction parameters are estimated from the size of the magnetic moments in the system.
Thus, the resulting interaction parameters tend to be large for systems with large spin-polarization.
\end{thebibliography}
\end{document}